\documentclass[twocolumn]{aastex63}
\shorttitle{Intermediate-Mass Black Holes}
\shortauthors{Wrobel et al.}

\begin{document}

\title{Accessing Intermediate-Mass Black Holes in 728 Globular Star
  Clusters in NGC\,4472}

\author[0000-0001-9720-7398]{J. M. Wrobel}
\affiliation{National Radio Astronomy Observatory, P.O. Box O,
  Socorro, NM 87801, USA}

\author[0000-0003-0976-4755]{T. J. Maccarone}
\affiliation{Department of Physics and Astronomy, Texas Tech
  University, Box 41051, Lubbock, TX 79409-1051, USA}

\author[0000-0003-3124-2814]{J. C. A. Miller-Jones}
\affiliation{International Centre for Radio Astronomy Research, Curtin
  University, GPO Box U1987, Perth, WA 6845, Australia}

\author[0000-0003-1991-370X]{K. E. Nyland}
\affiliation{National Research Council, Resident at the U.S. Naval
  Research Laboratory, 4555 Overlook Avenue SW, Washington, DC 20375,
  USA}

\correspondingauthor{J. M. Wrobel}
\email{jwrobel@nrao.edu}

\accepted{2021 June 25}
\submitjournal{ApJ}

\begin{abstract}
Intermediate-mass black holes (IMBHs) by definition have masses of
$M_{\rm IMBH} \sim 10^{2-5}~M_\odot$, a range with few observational
constraints. Finding IMBHs in globular star clusters (GCs) would
validate a formation channel for massive black-hole seeds in the early
universe. Here, we simulate a 60-hour observation with the
next-generation Very Large Array (ngVLA) of 728 GC candidates in the
Virgo Cluster galaxy NGC\,4472. Interpreting the radio detection
thresholds as signatures of accretion onto IMBHs, we benchmark IMBH
mass thresholds in three scenarios and find the following: (1) Radio
analogs of ESO\,243-49 HLX-1, a strong IMBH candidate with $M_{\rm
  IMBH}^{\rm HLX} \sim 10^{4-5}~M_\odot$ in a star cluster, are easy
to access in all 728 GC candidates. (2) For the 30 GC candidates with
extant X-ray detections, the empirical fundamental-plane relation
involving black hole mass plus X-ray and radio luminosities suggests
access to $M_{\rm IMBH}^{\rm FP} \sim 10^{1.7-3.6}~M_\odot$, with an
uncertainty of 0.44 dex. (3) A fiducial Bondi accretion model was
applied to all 728 GC candidates and to radio stacks of GC
candidates. This model suggests access to IMBH masses, with a
  statistical uncertainty of 0.39 dex, of $M_{\rm IMBH}^{\rm B} \sim
10^{4.9-5.1}~M_\odot$ for individual GC candidates and $M_{\rm
  IMBH}^{\rm B,stack} \sim 10^{4.5}~M_\odot$ for radio stacks of about
100-200 GC candidates.  The fiducial Bondi model offers initial
  guidance, but is subject to additional systematic uncertainties and
should be superseded by hydrodynamical simulations of gas flows in
GCs.
\end{abstract}


\section{Motivation}\label{1}

By definition, intermediate-mass black holes (IMBHs) have masses of
$M_{\rm IMBH} \sim 10^2 - 10^5~M_\odot$. Discovering them in
present-day globular star clusters (GCs) would validate a formation
channel for the seeds of massive black holes (BHs) in the early
universe \citep[for a review, see][]{gre20}. It would also inform
predictions for gravitational wave and tidal disruption events offset
from galactic nuclei \citep[e.g.,][]{fra18}.

To search for IMBHs in GCs, one looks for evidence that the IMBHs are
influencing the properties of their GC hosts. Within the Local Group,
a common approach is to scrutinize optical or infrared data for the
dynamical signatures of IMBHs on the orbits of stars in the
GCs. Further afield, the extremely large telescopes in the 30m class
aim to measure, at a distance of 10~Mpc, an IMBH mass of
$10^5~M_\odot$ by spatially resolving its sphere of influence in its
GC host \citep{do14}. Such sphere-of-influence searches are
susceptible to measuring high concentrations of stellar remnants
rather than an IMBH \citep[e.g.,][]{and10,van10,bau17,rui21}. It is
thus important to develop alternate ways to search for IMBHs in GCs.

One alternate approach, first suggested by \citet{mac04}, builds on
extensive studies of accretion signatures from stellar-mass and
supermassive BHs. By analogy with stellar-mass BHs \citep[reviewed
  by][]{fen12}, it is expected that an IMBH will spend more time in
the hard X-ray state - including quiesence - than in the soft X-ray
state.  In the typical case of just a few radio observations, it is
likely that they will sample the steady radio emission characterizing
the hard X-ray state, rather than the flaring radio emssion occuring
during a transition from the hard X-ray state to the soft X-ray
state. These concepts lead to the following three scenarios:

\begin{itemize}
\item Seek radio emission resembling that from ESO\,243-49 HLX-1 in
  its hard X-ray state. HLX-1 is a strong IMBH candidate in a star
  cluster which may have a GC-like stellar mass
  \citep{far09,sor12,sor17,cse15}.
\item Use the empirical fundamental-plane regression for the hard
  X-ray state, plus observations of X-ray and radio luminosities, to
  estimate an IMBH mass \citep{mer03,fal04,plo12,mil12}.
\item Use a fiducial, semi-empirical model to predict the mass of an
  IMBH that, if experiencing Bondi accretion in the hard X-ray state,
  would be consistent with the observed radio luminosity
  \citep{mac08,mac10,str12}.
\end{itemize}

It is expected that the radio continuum emission from an IMBH will be
persistent over time, flat-spectrum, jet-like but spatially unresolved
at the anticipated resolutions, and located at or close to the
dynamical center of the GC \citep{mac04}. Where necesary, we adopt a
radio spectral index of zero when evaluating radio luminosities or
comparing observations at different wavelengths.

Evolutionary models of GCs show that not all are able to retain their
IMBHs against gravitational wave or Newtonian recoils
\citep[e.g.,][]{hol08,fra18}. Predictions for the overall retention
fraction for a galaxy's GC population are quite model dependent, but
could be as low as a few percent \citep{fra18}. Examining a large
number of GCs per galaxy is thus key, driving us to consider massive,
bulge-dominated galaxies \citep{har13}. Section~2 describes how we
select a galaxy and simulate a radio observation of it. In Section~3
we quantify the radio detection thresholds, interpret them as
signatures of accretion onto IMBHs, and benchmark the associated IMBH
mass thresholds within the context of the scenarios above. We close in
Section~4 with a summary and conclusions.

\section{Simulated Observation}\label{2}

Some massive, bulge-dominated galaxies exhibit strong radio continuum
emission \citep{nag05,nyl16}, potentially limiting the depths of radio
searches. We avoid M87 and other such galaxies, and focus on
NGC\,4472, at a distance of 16.7\,Mpc and the optically brightest
galaxy in the Virgo Cluster \citep{bla09}. Ground-based imaging of
NGC\,4472 over tens of arcminutes suggests that the total number of
GCs is about 9000, with their surface density peaking in the innermost
few arcminutes \citep{dur14}. Here we focus on that innermost region
of NGC\,4472 and adopt the \citet{mac03} catalog of GC candidates
derived from deep HST/WFPC2 photometry in V(F555W) and I(F814W)
bands. We simulate an observation of those candidates with the
next-generation Very Large Array \citep[ngVLA;][]{sel18}.

A typical GC has a half-starlight diameter of 5\,pc \citep{bro06},
which subtends 60\,mas at NGC\,4472's distance. We thus select a
wavelength of 2cm to ensure an adequately fine spatial resolution of
100\,mas (8.1\,pc) and an adequately large field of view (FoV) to
encompass hundreds of GC candidates in one ngVLA pointing. Balancing
loss of sensitivity due to primary beam attenuation against the number
of GC candidates, we adopt a FoV of 5.09\,arcmin (24.7\,kpc) set by
the primary beam at full width at quarter maximum (FWQM).

GC candidates were identified via their size and color \citep{mac03}.
We estimate the stellar masses $M_\star$ of the GC candidates from the
I-band photometry, assuming a Solar value of 4.52 AB mag \citep{wil18}
and a mass-to-light ratio of 1.4 \citep{jor07}. After excluding one GC
candidate as likely being too massive for a bona fide GC
\citep{nor19}, we are left with 728 GC candidates in a single ngVLA
pointing (Figure~1). As yet, there is no information as to which GC
candidates are core-collapsed and which are not. The consensus of
theoretical studies is that core-collapsed GCs cannot harbour IMBHs
\citep[][and references therein]{rui21}. The ngVLA pointing will
encompass both types.

\begin{figure}[t!]
\epsscale{1.3}
\plotone{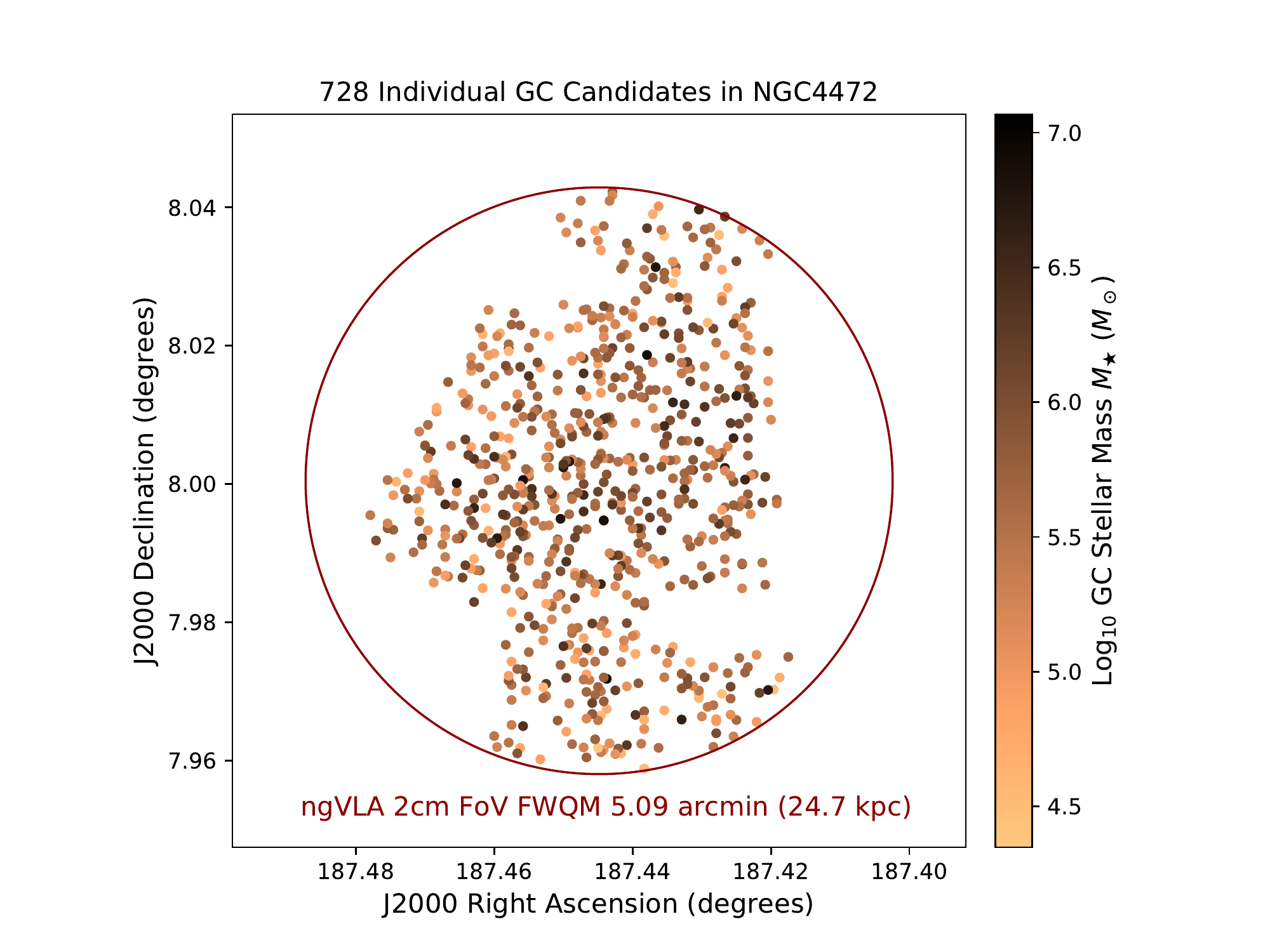}  
\caption{Locations of 728 individual GC candidates in NGC\,4472
  \citep{mac03} and within a FoV equal to FWQM of the primary beam of
  an ngVLA antenna at a wavelength of 2cm \citep{sel18}. The color
  gradation conveys estimates of the stellar masses $M_\star$ of the
  GC candidates.}
\end{figure}

We then pick a reasonable observation duration on NGC\,4472 of
60\,hours. Using the ngVLA Sensitivity Calculator
(https://gitlab.nrao.edu/vrosero/ngvla-sensitivity-calculator), the
Main Array of 214 18-m antennas at a wavelength of 2cm will provide a
detection threshold of $S_{\rm 2cm} = 3 \times 0.034~\mu$Jy
beam$^{-1}$ at 3$\sigma$ at the peak of the primary beam. NGC\,4472
hosts a low-luminosity AGN with a peak 2cm flux density of 3.7~mJy for
an angular resolution of 150\,mas \citep{nag05}. Imaging NGC\,4472
with the ngVLA will thus require a dynamic range of about
$(3700/0.034) \sim 10^5$, achievable based on the facility's planned
performance \citep{sel18}. As \citet{nyl18} emphasize, those
contemplating deep ngVLA searches -- as simulated here -- should be
aware that sky regions containing only a few mJy can start to suffer
from dynamic-range limitations.

\section{Implications}\label{3}

\subsection{Radio Luminosity Thresholds}\label{3.1}

At the location of each GC candidate, we correct the radio detection
threshold for the primary beam attenuation at 2cm and then convert the
corrected detection threshold to a radio luminosity threshold $L_{\rm
  2cm}$ at NGC\,4472's distance. As Figure~2 demonstrates, the radio
luminosity thresholds reach $L_{\rm 2cm} \sim 10^{32.7-33.3}$
erg~s$^{-1}$.

\begin{figure}[t!]
\epsscale{1.2}
\plotone{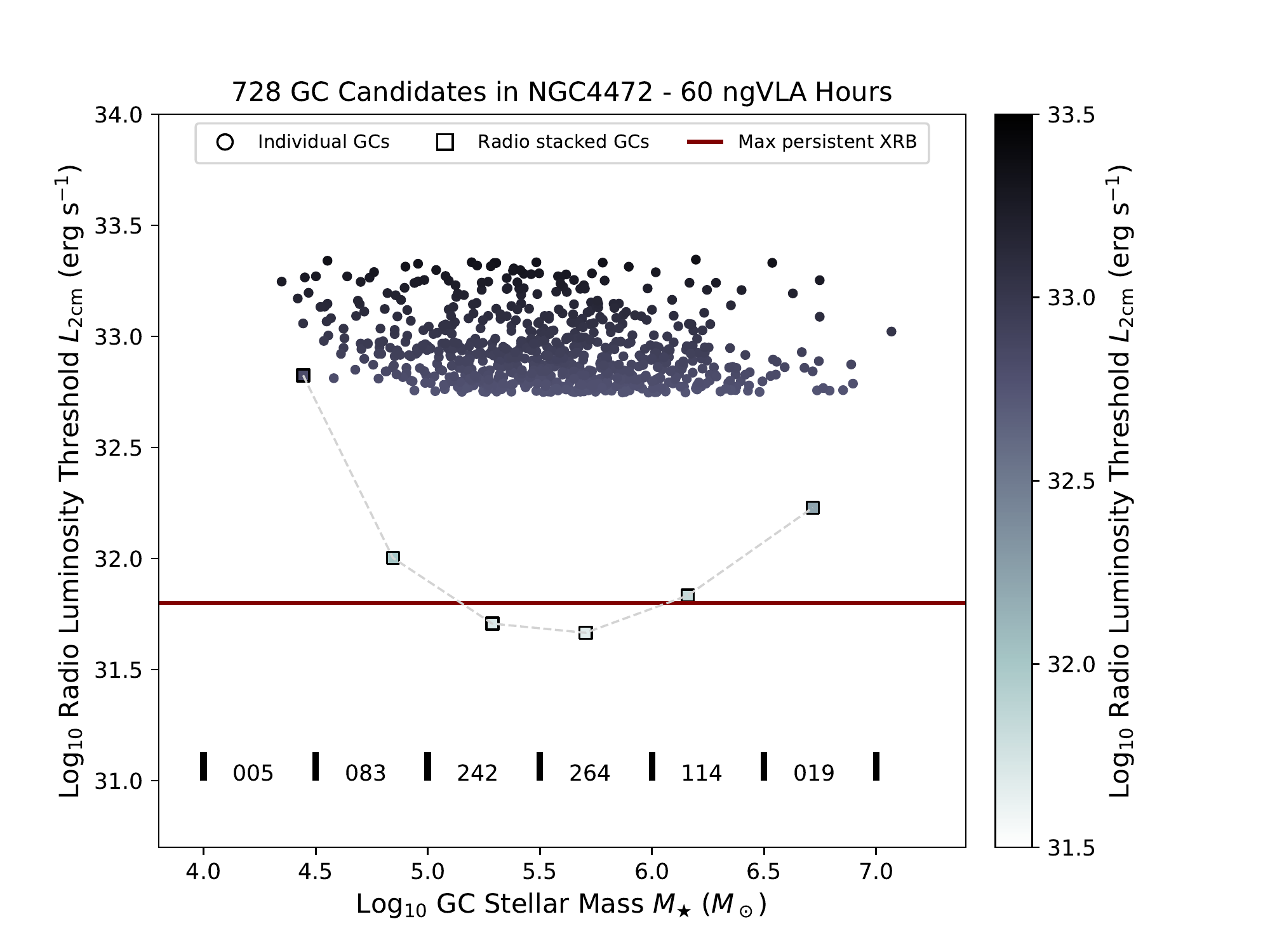}
\caption{Radio luminosity thresholds $L_{\rm 2cm}$ at 3$\sigma$ from a
  simulated ngVLA observation of GC candidates in NGC\,4472, versus
  the stellar masses $M_\star$ of the GC candidates. The color
  gradation encodes the radio luminosity thresholds. The circles mark
  728 individual GC candidates \citep{mac03}. The spread in $L_{\rm
    2cm}$ arises from the primary beam attenuation at the locations of
  the GC candidates. The numbers of GC candidates in bins of width
  0.5\,dex in stellar mass $M_\star$ are noted. The squares mark the
  radio luminosity thresholds $L_{\rm 2cm}^{\rm stack}$ after stacking
  the GC candidates in each of the bins. The horizontal line shows the
  maximum persistent luminosity of stellar-mass XRBs \citep{bah20}.}
\end{figure}

To explore the prospects for deeper detections, we also wish to stack
the corrected radio detection thresholds of GC candidates with similar
stellar masses. Techniques for stacking extragalactic GC populations
have been developed using VLA images of nearby galaxies
\citep{wro15,wro16,wro20}. Here, we opt to segregate NGC\,4472's
individual GC candidates into bins of width 0.5\,dex in stellar mass
$M_\star$.  For each bin, we evaluate its number of GC candidates,
median $M_{\star}^{\rm stack}$, weighted-mean detection threshold
$S_{\rm 2cm}^{\rm stack}$, and associated radio luminosity threshold
$L_{\rm 2cm}^{\rm stack}$ (Table~1). We caution that such stacking
implicitly assumes a high occupation fraction of IMBHs in each bin's
GCs. Meeting this condition might be easier in the high-stellar-mass
bins: GC evolutionary models suggest that the more massive a GC, the
more likely it is to retain a putative IMBH against gravitational wave
or Newtonian recoils \citep[e.g.,][]{hol08,fra18}.

Regarding possible radio contaminants behind NGC\,4472's GC
candidates, simulated source counts near 2cm suggest that star forming
galaxies will dominate at $\mu$Jy levels \citep{wil08}. As such
galaxies have steep specta and finite sizes, they will not be mistaken
for IMBHs that have flat spectra and are point-like.

Regarding possible radio contaminants within NGC\,4472's GC
candidates, studies of stellar-mass X-ray binaries (XRBs) in the Milky
Way show that their persistent radio luminosities at 6cm top out at
about $10^{31.3}$ erg~s$^{-1}$ for BH systems \citep{bah20}. Scaling a
flat-spectrum contaminant from 6cm to 2cm in an individual GC, this
translates to a radio luminosity $L_{\rm 2cm} \sim 10^{31.8}$
erg~s$^{-1}$, well below the radio luminosity thresholds of individual
GC candidates in NGC\,4472 (Figure~2). If every GC harboured such an
extreme contaminant, a signal could just emerge above the deepest
stacked thresholds $L_{\rm 2cm}^{\rm stack}$ (Figure~2, Table~1); we
dismiss such a contrived situation. Also, one radio-flaring XRB in an
ultraluminous state has been identified in M31 \citep{mid13}; while a
radio analog of it could be detected in NGC\,4472, it would fade after
a few months so not be mistaken for the persistent emitters that we
seek.

\subsection{X-ray Detections}\label{3.2}

Only 30 of the 728 GC candidates, or 4\%, have been detected with
Chandra \citep{mac03}. Their X-ray luminosities $L_{\rm X}^{\rm det}$
are shown in Figure~3 after being adjusted to our assumed distance for
NGC\,4472. If these 30 detections arise from BH XRBs, their
luminosities would correspond to systems in their high/soft state or
very high state. However, the signal-to-noise of the X-ray
spectroscopy was insufficient to establish the emission states. Here,
we assume that the 30 detections arise from BHs of unspecified mass
that are emitting in the hard X-ray state. If improved X-ray
spectroscopy eventually excludes hard-state emission, the $L_{\rm
  X}^{\rm det}$ values should be treated as upper limits on any
hard-state emission.

\begin{figure}[t!]
\epsscale{1.2}
\plotone{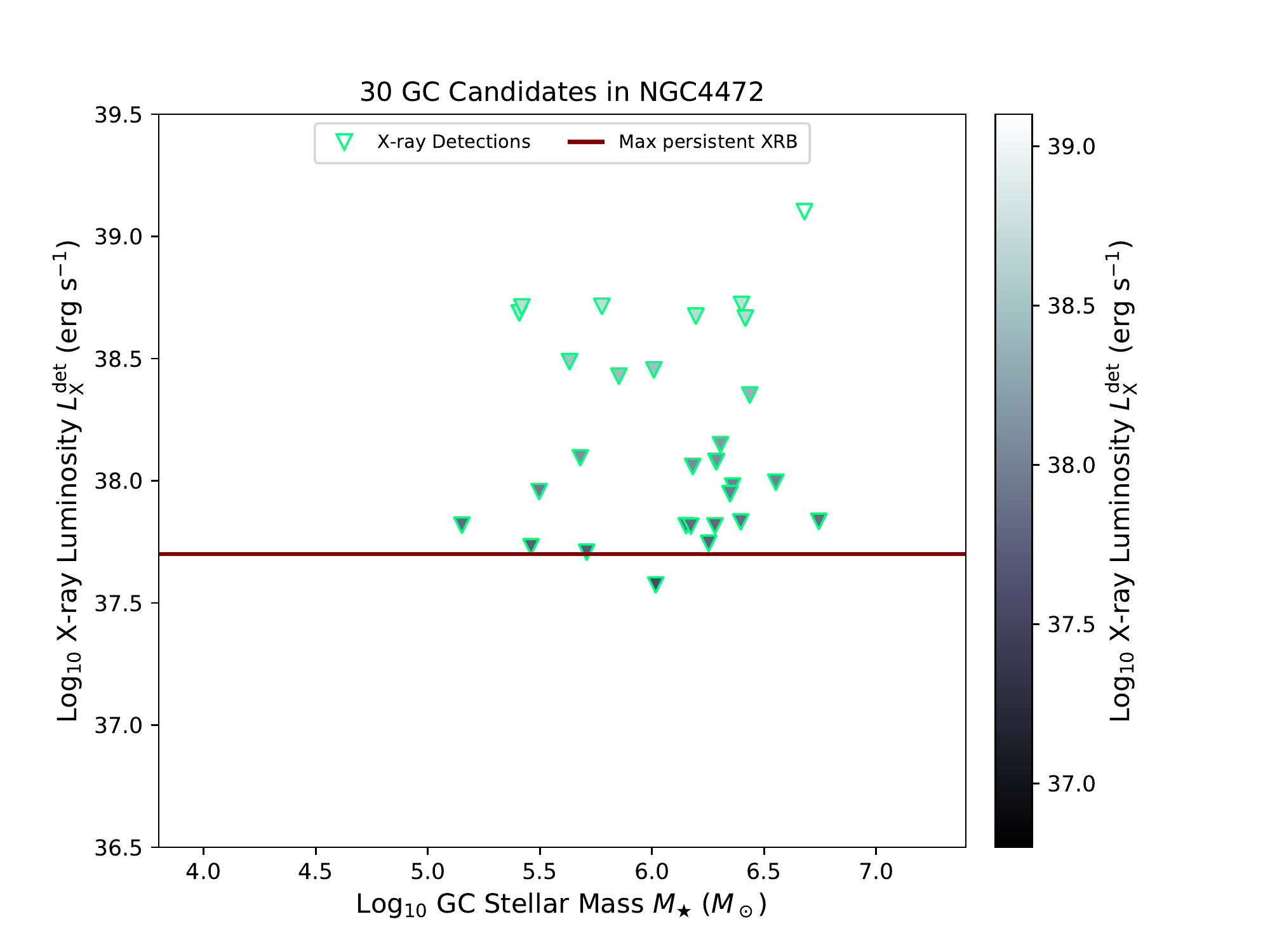}
\caption{X-ray luminosities $L_{\rm X}^{\rm det}$ of 30
  Chandra-detected GC candidates in NGC\,4472 \citep{mac03} versus
  their stellar masses $M_\star$. The color gradation encodes the
  X-ray luminosities. The horizontal line shows the maximum persistent
  luminosity of stellar-mass X-ray binaries \citep{mac03}.}
\end{figure}

\subsection{Analogs of ESO\,243-49 HLX-1}\label{3.3}

ESO\,243-49 HLX-1 is a strong IMBH candidate, of mass $M_{\rm
  IMBH}^{\rm HLX} \sim 10^{4-5}~M_\odot$, in a star cluster which may
have a GC-like stellar mass \citep{far09,sor12,sor17}. 
  \citet{cse15} conclude that a point-like source with a flux density
  of $22 \pm 3.5~\mu$Jy at 6.8\,GHz adequately describes the
  emission. For the flat spectrum appropriate to the hard X-ray state,
  the same flux density is adopted at 2cm (16.4\,GHz). Thus for a
luminosity distance of 92\,Mpc, HLX-1 is expected to have a steady 2cm
luminosity of $L_{\rm 2cm} \sim 10^{36.6}$ erg~s$^{-1}$ while in its
hard X-ray state \citep{cse15}. If that steady emission is being
Doppler boosted by a factor of about five to ten, as \citet{cse15}
argue, then its side-on luminosity would be about $L_{\rm 2cm} \sim
10^{35.6-35.8}$ erg~s$^{-1}$. Given these various radio luminosities,
Figure~2 makes it clear that radio analogs of HLX-1 would be easy to
access among NGC\,4472's 728 GC candidates.

Figure~3 shows that none of NGC\,4472's GC candidates has an observed
X-ray luminosity \citep{mac03} as high as that of HLX-1 in its hard
state, $L_{\rm X} \sim 10^{40.3}$ erg~s$^{-1}$ \citep{god12}. This
absence of X-ray analogs of HLX-1 would be consistent with an absence
of radio analogs of HLX-1, discussed above. However, the radio
luminosity thresholds would be more than two orders of magnitude lower
than even a deboosted version of HLX-1. This would enable searches for
less radio-extreme sources, even in the case where their X-ray
counterparts were heavily absorbed, such that their X-ray emission was
not especially remarkable.

\subsection{IMBH Masses and the Fundamental Plane}\label{3.4}

Figure~3 shows the X-ray luminosities $L_{\rm X}^{\rm det}$ for 30 GC
candidates in NGC\,4472 \citep{mac03}. As Section~3.2 mentions, we
assume that those 30 X-ray detections arise from BHs of unspecified
mass that are emitting in the hard X-ray state. Radio luminosity
thresholds $L_{\rm 2cm}$ are also available from Figure~2 for those 30
GC candidates.

\begin{figure}[t!]
\epsscale{1.2}
\plotone{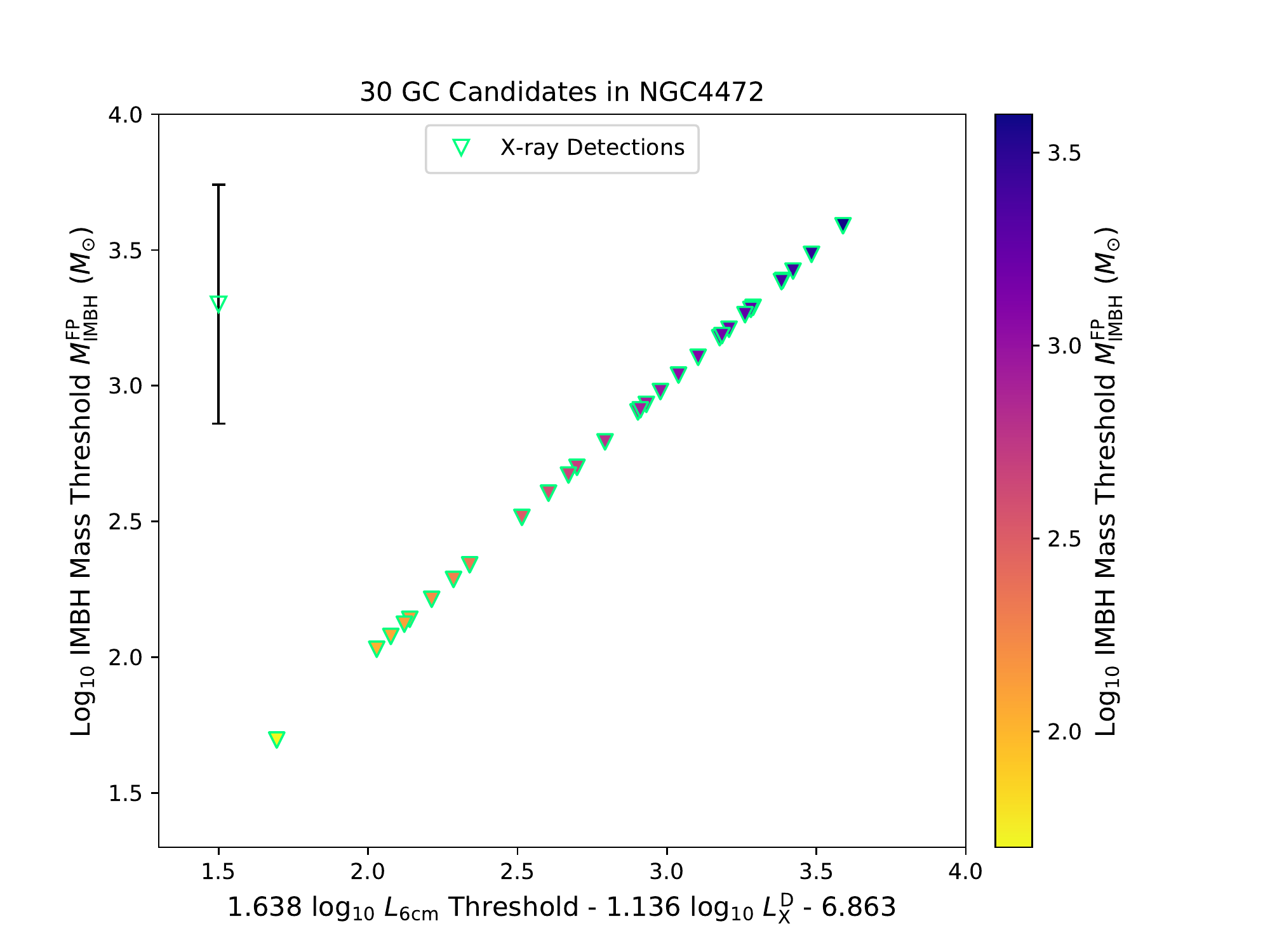}
\caption{IMBH mass thresholds suggested from the fundamental plane,
  $M_{\rm IMBH}^{\rm FP}$, for the simulated ngVLA observation of 30
  GC candidates in NGC\,4472 with Chandra detections \citep{mac03}.
  The color gradation encodes $M_{\rm IMBH}^{\rm FP}$ and the example
  error bar shows its 0.44\,dex uncertainty \citep{mil12}.}
\end{figure}

Armed with this information for the 30 GC candidates, we can invoke
the empirical fundamental-plane (FP) relation for the hard X-ray state
to estimate their IMBH mass thresholds $M_{\rm IMBH}^{\rm FP}$, with
an uncertainty of 0.44\,dex \citep{mil12}. That FP relation adopts an
X-ray range of 0.5-10\,keV and a radio wavelength of 6cm, and
  implicitly assumes a flat or inverted radio spectrum. The $L_{\rm
  X}^{\rm det}$ values from \citet{mac03} cover 0.5-8\,keV which, at
the signal-to-noise ratios involved, is a suitable proxy for the FP's
range. The $L_{\rm 2cm}$ thresholds from Figure~2 were scaled to
$L_{\rm 6cm}$ assuming a flat radio spectrum. While other FP studies
\citep[e.g.,][]{gul19} suggest larger uncertainties than those
advocated by \citet{mil12}, we follow \citet{mil12} for consistency
with our earlier work on Milky Way GCs \citep{mac04,str12} and
extragalactic GCs \citep{wro15,wro16,wro18,wro20}.

The FP results are shown in Figure~4. They demonstrate that
measurements of $L_{\rm X}^{\rm det}$ and thresholds for $L_{\rm 6cm}$
provide access to IMBH mass thresholds of $M_{\rm IMBH}^{\rm FP} \sim
10^{1.7-3.6}~M_\odot$. Notably, among previous attempts to apply the
FP to extragalactic GCs \citep{mil12,wro15,wro16}, this regime has
only been reached in the Local Group, for M31's G1 with its $M_{\rm
  IMBH}^{\rm FP} < 10^{3.2}~M_\odot$ \citep{mil12}.

Each IMBH mass threshold in Figure~4 suffers from a considerable
scatter of 0.44\,dex. Still, such thresholds do open up the prospect
of searching for accretion signatures from IMBHs in tens of GC
candidates at a distance of 16.7\,Mpc, in the GC-rich Virgo
Cluster. Deeper X-ray imaging of NGC\,4472, with either Chandra or its
possible successor, Lynx, could reveal sources with lower values of
$L_{\rm X}^{\rm det}$ in its 698 other GC candidates. Each such new
datum added to Figure~4 would have a rightward shift on the abscissa
and an upward shift on the ordinate, thereby giving access to a higher
$M_{\rm IMBH}^{\rm FP}$ value.

\subsection{IMBH Masses and Bondi Accretion}\label{3.5}

This scenario involves predicting the mass of an IMBH that, if
accreting at the Bondi rate from the tenuous gas supplied to the GC
from its evolving stars, is consistent with the synchrotron radio
luminosity of the GC \citep{mac04,str12}. Classic Bondi flows involve
perfect gases, with spatially infinite distributions, accreting onto
isolated central masses, IMBHs in our case. This scenario may be
overly simplistic for GC settings but can still offer useful guidance.

Some systematic uncertainties need to be recognized. First, the
thermodynamic state -- isothermal, adiabatic or intermediate between
the two -- of the gas flow feeding the Bondi accretion is unknown in
GCs. To quantify the importance of this shortcoming, we note that the
accretion rate for the isothermal Bondi case is almost a factor of ten
higher than for the adiabatic case \citep[e.g.,][]{pel05}. For a given
IMBH mass, the accretion signature from a Bondi flow will thus be
stronger for the isothermal case.

However, the study by \citet{pep13} raises a second systematic
uncertainty: for four Milky Way GCs that study numerically solved the
accretion rates for isothermal Bondi flows toward IMBHs embedded in
realistic GC potentials. They reported that the flows appeared to
achieve higher accretion rates than classic isothermal Bondi flows.

Recent hydrodynamical simulations of gas flows in early-type galaxies
favor an adiabatic behavior \citep{ina20}, with its lower accretion
rates in comparison to the isothermal case \citep{pel05}. Future
hydrodynamical simulations of gas flows in GCs will eventually
establish their true physical properties. In the interim, our aim here
is to simply benchmark the accretion signatures. Being mindful of the
systematic uncertainties mentioned above, we follow our earlier work
on Milky Way GCs \citep{mac04,str12} and extragalactic GCs
\citep{wro15,wro16,wro18,wro20}, and assume fiducial Bondi flows where
the thermodynamic state of the gas is intermediate between the
isothermal and adiabatic cases. Specifically, we assume \citep{str12}:

\begin{enumerate}
\item[$i.$] For a given IMBH mass $M_{\rm IMBH}^{\rm B}$, gas is
  captured at 3\% of a fiducial Bondi rate for a density of 0.2
  particles cm$^{-1}$ \citep{fre01,abb18} and a temperature of
  $10^4$~K \citep{sco75}. We caution that the density estimate hinges
  on measurements of only two objects, namely the Milky Way GCs
  47\,Tuc and M15.
\item[$ii.$] An X-ray luminosity given by $L_{\rm X}^{\rm B} =
  \epsilon \dot{M} c^2$, for a radiative efficiency $\epsilon$ and
  mass accretion rate $\dot{M}$. To generalize beyond the standard
  $\epsilon_{\rm sta} = 0.1$, it is assumed that at low accretion
  rates the efficiency scales linearly with the accretion rate. A low
  accretion rate is defined as being less than 2\% of the Eddington
  rate, typical for where XRBs transition to a state with an inner
  advection-dominated accretion flow with a persistent X-ray
  luminosity. Continuity with $\epsilon_{\rm sta} = 0.1$ is required
  at the transition, leading to $\epsilon = 0.1 ((\dot{M}/\dot{M_{\rm
      edd}}) / 0.02)$ and thus a prediction for $L_{\rm X}^{\rm B}$.
\item[$iii.$] Given $M_{\rm IMBH}^{\rm B}$ and $L_{\rm X}^{\rm B}$, a
  radio luminosity $L_{\rm 6cm}$ can be estimated from the empirical
  fundamental-plane relation. A flat radio spectrum is used to scale
  from $L_{\rm 6cm}$ to $L_{\rm 2cm}$.
\end{enumerate}

With assumptions $i-iii$, the radio luminosity thresholds $L_{\rm
  2cm}$ shown in Figure~2 can be converted to the IMBH mass thresholds
$M_{\rm IMBH}^{\rm B}$ displayed in Figure~5, for both individual GC
candidates and radio-stacked GC candidates. The statistical
uncertainties in the parameters adopted for the conversion cause each
IMBH mass threshold to have a statistical error of 0.39\,dex
\citep{str12}. The fiducial Bondi model suggests access to IMBH masses
of $M_{\rm IMBH}^{\rm B} \sim 10^{4.9-5.1}~M_\odot$ for all 728
individual GC candidates and of $M_{\rm IMBH}^{\rm B,stack} \sim
10^{4.5}~M_\odot$ for radio stacks of about 100-200 GC candidates
(Table~1). Encouragingly, that stack level at a distance of 16.7\,Mpc
resembles the stack level reported from VLA observations of 49 massive
GCs in M81 at a distance of 3.6\,Mpc \citep{wro16}.

\begin{figure}[t!]
\epsscale{1.2}
\plotone{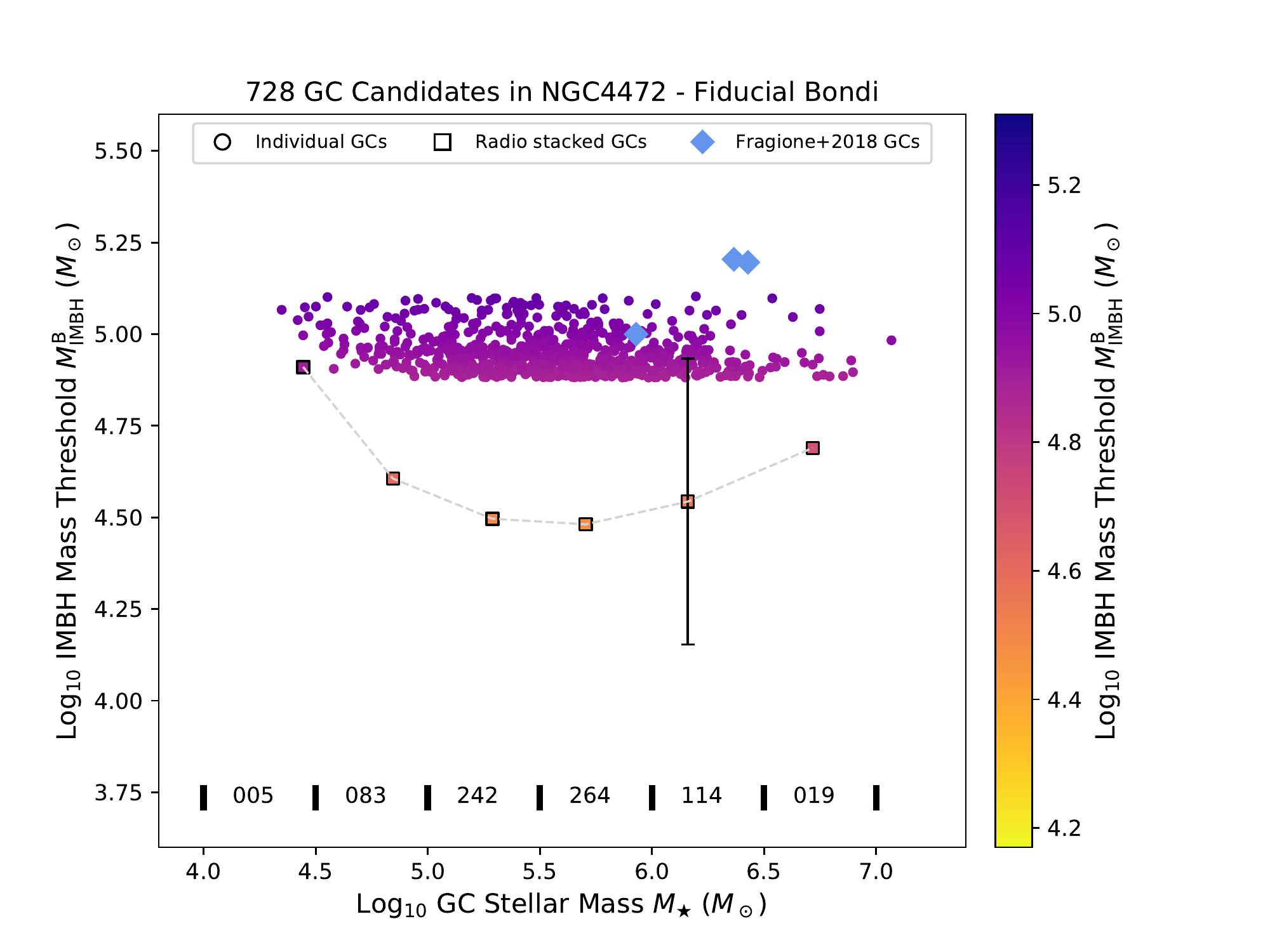}
\caption{IMBH mass thresholds $M_{\rm IMBH}^{\rm B}$ predicted from a
  fiducial Bondi accretion model of GC candidates in NGC\,4472
  \citep{mac03}, versus the stellar masses $M_\star$ of the GC
  candidates. The color gradation encodes the IMBH mass thresholds.
  The circles mark 728 individual GC candidates. Their spread in
  $M_{\rm IMBH}^{\rm B}$ reflects the spread in $L_{\rm 2cm}$ in
  Figure~2. The numbers of GC candidates in bins of width 0.5\,dex in
  stellar mass $M_\star$ are noted. The squares mark the IMBH mass
  thresholds $M_{\rm IMBH}^{\rm B,stack}$ after stacking the GC
  candidates in each of the bins. The error bar shows the 
    statistical uncertainty of 0.39\,dex in each IMBH mass threshold,
  whether individual or stacked. Example predictions from a
  semi-analytic model for GC evolution over cosmic time are shown
  \citep{fra18}.}
\end{figure}

For context, \citet{fra18} offer three examples of how a star
cluster's stellar mass and IMBH mass could evolve to yield a massive,
present-day GC. Those examples, plotted in Figure~5, achieve
present-day values of $M_\star^{\rm Fra} \sim 10^{5.9-6.4}~M_\odot$
and $M_{\rm IMBH}^{\rm Fra} \sim 10^{5.0-5.2}~M_\odot$. At similar
stellar masses, the $M_{\rm IMBH}^{\rm Fra}$ values are formally
within twice the statistical error on the $M_{\rm IMBH}^{\rm B}$
values, whether for individual GCs or radio-stacked GCs. But this is a
weak statement, due to the substantial statistical uncertainties of
the fiducial Bondi model as well as to the aforementioned systematic
uncertainties.

\begin{figure}[t!]
\epsscale{1.2}
\plotone{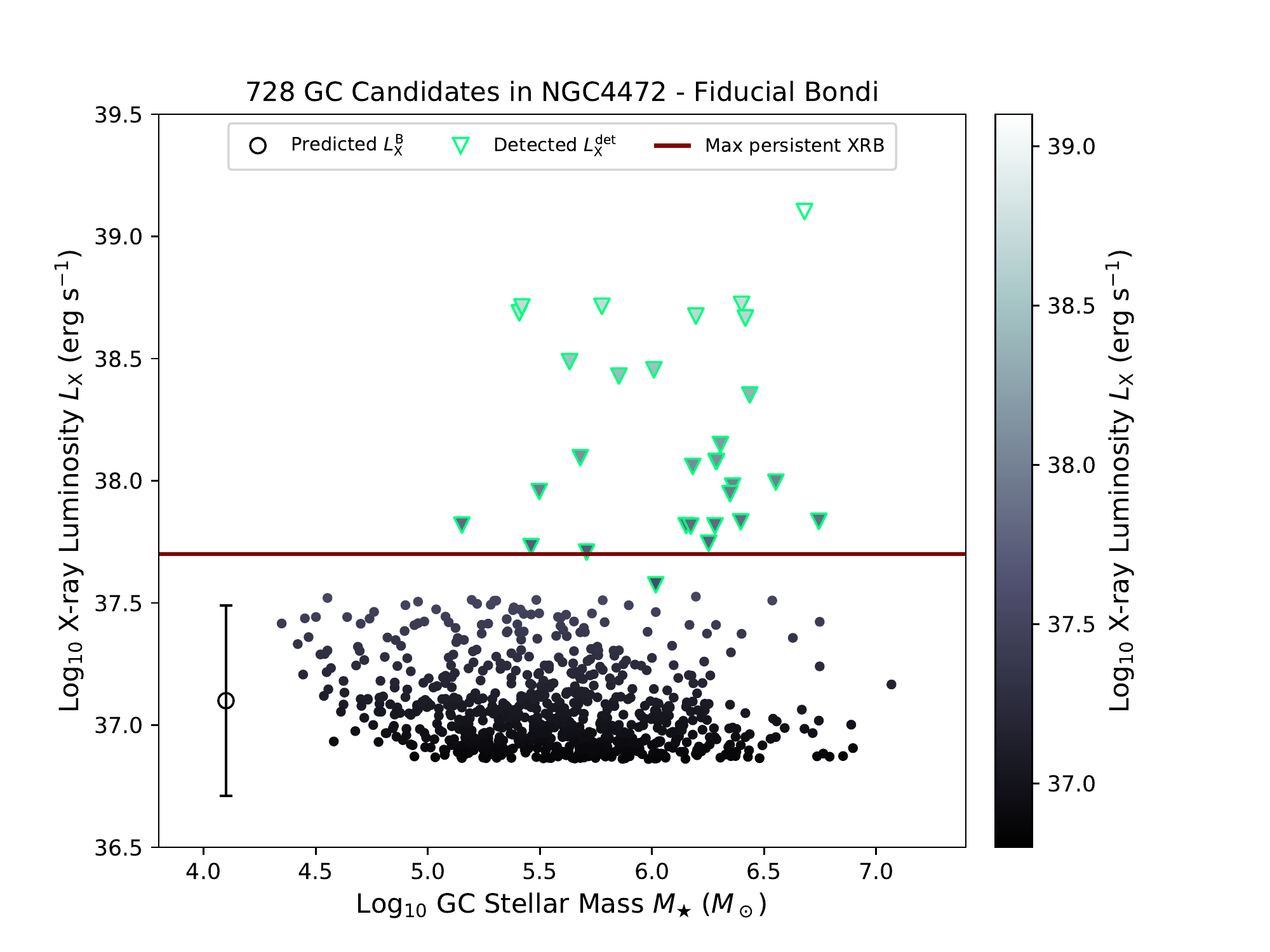}
\caption{Circles show the X-ray luminosity thresholds $L_{\rm X}^{\rm
    B}$ predicted from a fiducial Bondi accretion model for 728 GC
  candidates in NGC\,4472 \citep{mac03}, versus the stellar masses
  $M_\star$ of the GC candidates. The error bar shows the statistical
  uncertainty of 0.39\,dex in each $L_{\rm X}^{\rm B}$ value, with the
  spread among values reflecting the spread in $L_{\rm 2cm}$ in
  Figure~2. The triangles show the X-ray luminosities $L_{\rm X}^{\rm
    det}$ of the 30 Chandra-detected GC candidates in Figure~3. The
  color gradation encodes the X-ray luminosities. The horizontal line
  shows the maximum persistent luminosity of stellar-mass XRBs
  \citep{mac03}.}
\end{figure}

For the 728 GC candidates, the fiducial Bondi model also yields
predictions of $L_{\rm X}^{\rm B}$, the X-ray luminosity thresholds
for hard-state emission in the 0.5-10\,keV band. These predictions are
shown in Figure~6, along with an indication of their statistical
uncertainty of 0.39\,dex \citep{str12}. The $L_{\rm X}^{\rm B}$
thresholds appear to cluster below the $L_{\rm X}^{\rm det}$ values of
the 30 Chandra-detected GC candidates. As discussed in Section~3.2,
those 30 X-ray detections could correspond to XRB systems in their
high/soft state or very high state, with no relation at all to the
presence or absence of accreting IMBHs. But if some of the 30 X-ray
detections are hard-state emitters involving accreting IMBHs, Figure~6
suggests that the fiducial Bondi model could be underpredicting their
$L_{\rm X}^{\rm B}$ thresholds.

Figure~6 also indicates that the predicted $L_{\rm X}^{\rm B}$
thresholds for all 728 GC candidates appear to cluster below the
maximum persistent luminosity of stellar-mass XRBs \citep{mac03}.
While deeper X-ray imaging could reveal additional sources with lower
values of $L_{\rm X}^{\rm det}$ in the GC candidates, such imaging on
its own would not be able to say if any detections of hard-state X-ray
emission arose from stellar-mass XRBs or from fiducial Bondi flows
onto IMBHs. This represents a serious X-ray contamination issue, and
is the reason why we opted to avoid reporting in Table~1 the X-ray
luminosities from the radio-stacked GCs.

We do not mean to discourage deeper X-ray imaging of NGC\,4472 with
either Chandra or its possible successor, Lynx: such imaging, when
combined with ngVLA imaging and a FP analysis, can be used to separate
X-ray detections into bins for X-ray binaries and for IMBHs
\citep{wro18}. Also, X-ray binaries are known to be time-variable in
both the radio and X-ray bands, so such radio--X-ray synergy would be
strengthened by simultaneous observations with the ngVLA and the X-ray
mission.

\section{Summary and Conclusions}\label{4}

We simulated a 60-hour observation with the ngVLA of 728 GC candidates
in NGC\,4472, at a distance of 16.7\,Mpc and the optically brightest
galaxy in the Virgo Cluster. Interpreting the 2cm detection thresholds
as signatures of accretion onto IMBHs, we benchmarked IMBH mass
thresholds in three scenarios and found:

\begin{enumerate}
\item[1.] Radio analogs of ESO\,243-49 HLX-1, a strong IMBH candidate
  with $M_{\rm IMBH}^{\rm HLX} \sim 10^{4-5}~M_\odot$ in a star
  cluster, are easy to access in all GC candidates.
\item[2.] For the 30 GC candidates with existing X-ray detections, the
  empirical fundamental-plane relation suggested access to $M_{\rm
    IMBH}^{\rm FP} \sim 10^{1.7-3.6}~M_\odot$, with an uncertainty of
  0.44\,dex. Such IMBH levels have been reached only for M31's G1 at a
  distance of less than 1\,Mpc.
\item[3.] For all GC candidates and for radio stacks of about 100-200
  GC candidates, a fiducal Bondi accretion model suggested access to
  IMBH masses of $M_{\rm IMBH}^{\rm B} \sim 10^{4.9-5.1}~M_\odot$ and
  $M_{\rm IMBH}^{\rm B,stack} \sim 10^{4.5}~M_\odot$, with 
    statistical uncertainties of 0.39\,dex. That IMBH stack level
  resembles the stack level reported for GCs in M81 at a distance of
  3.6\,Mpc. The fiducial Bondi model offers initial guidance, but
    is subject to additional systematic uncertainties and should be
  superseded by hydrodynamical simulations of gas flows in GCs.
\end{enumerate}

\acknowledgments We thank the reviewer for a helpful and timely
  report. The NRAO is a facility of the National Science Foundation
(NSF), operated under cooperative agreement by AUI. The ngVLA is a
design and development project of the NSF operated under cooperative
agreement by AUI. Basic research in radio astronomy at the U.S. Naval
Research Laboratory is supported by 6.1 Base Funding.

\software{astropy (The Astropy Collaboration 2018)}

\begin{deluxetable*}{lcccccc}
  \tablecolumns{7} \tablewidth{0pc}
  \tablecaption{Stacked GC Candidates in NGC\,4472}
  \tablehead{\colhead{Stacked Attribute} &
  \colhead{Bin 1} &  \colhead{Bin 2} &  \colhead{Bin 3} &
  \colhead{Bin 4} &  \colhead{Bin 5} &  \colhead{Bin 6} }
\startdata
Number of GC candidates&      5&     83&    242&    264&    114&     19\\
Median GC Stellar Mass, Log$_{10}~M_\star^{\rm stack}$ ($M_{\odot}$)&
4.44&   4.84&   5.29&   5.70&   6.16&   6.72\\
Flux Density Threshold, $S_{\rm 2cm}^{\rm stack}$ ($\mu$Jy~beam$^{-1}$)&
0.122&  0.018&  0.009&  0.008&  0.012&  0.031\\
Radio Luminosity Threshold, Log$_{10}~L_{\rm 2cm}^{\rm stack}$ (erg~s$^{-1}$)&
32.82&  32.00&  31.71&  31.67&  31.83&  32.23\\
IMBH Mass Threshold, Fiducial Bondi, Log$_{10}~M_{\rm IMBH}^{\rm B,stack}$ ($M_\odot$)&
4.91&   4.61&   4.50&   4.48&   4.54&   4.69\\
\enddata
\end{deluxetable*}

\end{document}